\documentclass[runningheads]{llncs}
\usepackage{graphicx}
\usepackage[hidelinks]{hyperref}
\usepackage{tikz}
\usepackage{amsmath,bm}
\usepackage{dsfont}
\usepackage{siunitx}
\usepackage{algorithm2e}

\PassOptionsToPackage{hyphens}{url}\usepackage{hyperref}
\usepackage{amssymb}
\usepackage{todonotes}
\usepackage{float}

\usepackage{listings}

\definecolor{darkred}{RGB}{193, 39, 45}
\definecolor{indigo}{RGB}{0, 0, 167}
\definecolor{teal}{RGB}{0, 129, 118}
\definecolor{yellow}{RGB}{238, 204, 22}
\definecolor{lightgray}{RGB}{179, 179, 179}

\begin{document}
\title{Experiences Porting Shared and Distributed Applications to Asynchronous Tasks: A Multidimensional FFT Case-study}
\titlerunning{Experiences Porting Distributed Applications to Asynchronous Tasks}
\author{Alexander Strack\inst{1}\orcidID{0000-0002-9939-9044}  \and Christopher Taylor \inst{2}\orcidID{0000-0001-7119-818X} 
\and Patrick Diehl\inst{3}\orcidID{0000-0003-3922-8419} 
\and Dirk Pflüger\inst{1} \orcidID{0000-0002-4360-0212}}
\authorrunning{A. Strack \and et. al}
%
\institute{Institute of Parallel and Distributed Systems, University of Stuttgart,\\ 70569 Stuttgart, Germany\\ \email{\{alexander.strack, dirk.pflueger\}@ipvs.uni-stuttgart.de}
\and Tactical Computing Labs LLC, 1001 Pecan St., Lindsay, Texas, U.S.A.
\email{ctaylor@tactcomplabs.com}
\and 
Center of Computation and Technology, Louisiana State University, U.S.A. \email{patrickdiehl@lsu.edu}
}

\maketitle              
\begin{abstract}
Parallel algorithms relying on synchronous parallelization libraries often experience adverse performance due to global synchronization barriers. Asynchronous many-task runtimes offer task futurization capabilities that minimize or remove the need for global synchronization barriers. 
This paper conducts a case study of the multidimensional Fast Fourier Transform to identify which applications will benefit from the asynchronous many-task model. Our basis is the popular FFTW library \cite{Frigo2005}. We use the asynchronous many-task model HPX and a one-dimensional FFTW backend to implement multiple versions using different HPX features and highlight overheads and pitfalls during migration. Furthermore, we add an HPX threading backend to FFTW. 
The case study analyzes shared memory scaling properties between our HPX-based parallelization and FFTW with its pthreads, OpenMP, and HPX backends. The case study also compares FFTW's \textit{MPI+X} backend to a purely HPX-based distributed implementation. The FFT application does not profit from asynchronous task execution. In contrast, enforcing task synchronization results in better cache performance and thus better runtime. Nonetheless, the HPX backend for FFTW is competitive with existing backends. Our distributed HPX implementation based on HPX collectives using MPI parcelport has similar performance to FFTW's \textit{MPI+OpenMP}. However, the LCI parcelport of HPX accelerated communication up to factor $5$. 
\end{abstract}
\setcounter{footnote}{0} 
\section{Introduction}\label{sec:introduction}

Asynchronous many-task runtimes, supporting global address spaces, are a\newline promising alternative to message-based parallelization libraries like Open MPI \cite{Gabriel2004}. Synchronous parallelization libraries often require algorithms to use global synchronization barriers. However, applying global synchronization barriers can cause an algorithm to experience adverse performance. Task futurization features provided by asynchronous many-task runtimes can remove or minimize the use of global synchronization barriers.

Applications ported to asynchronous tasks, \emph{e.g.}, using HPX~\cite{hpx}, can result in notable performance gains. There is no free lunch meaning some applications are better suited than others for migration to an asynchronous task model. Applications solving inhomogeneous problems with few global synchronization barriers and no significant data exchange between different nodes can experience the introduction of undesirable overheads, \emph{e.g.}, by a global address space model and performance degradation.

We chose the multidimensional Fast Fourier Transform (FFT) as a case study to characterize applications better suited for migration to an asynchronous task model.
Many applications take advantage of the FFT, ranging from image compression~\cite{Wallace1991} to scientific applications like molecular dynamics~\cite{Deserno1998} and convolutional neural networks~\cite{Lavin2016}.
Discrete two-dimensional FFT is an excellent application for identifying overheads in a port to future-based, asynchronous tasks. The two-dimensional signal data is dense. A parallel algorithm partitions the data in a homogeneous and equal manner, which does not play to the strengths of HPX. The combination of global synchronization barriers and load imbalances will therefore make the task scheduler's ability to hide latency with work-stealing difficult.

The algorithm requires the exchange of data partitions between compute hosts, which are known in advance. The data partitions exchanged are homogeneous on all nodes and require node-level synchronization. The data transfer percentage is more extensive than in other applications, \emph{e.g.}, simulations in which small ghost layers must be communicated across neighboring nodes. Section~\ref{sec:methods} provides a more detailed explanation of the required synchronization and communication steps.

Our main contributions in this work include:
\begin{itemize}
    \item First distributed FFT proof-of-concept that solely relies on the HPX runtime model for distributed and shared memory parallelization
    \item Evaluation of the scaling properties of several implementation variants to identify overheads and highlight pitfalls when porting applications to futurized, asynchronous tasks
    \item Comparison against the de-facto standard, the FFTW library and its built-in parallelization backends 
    \item Implementation and performance comparison of an HPX backend for FFTW threading
\end{itemize}

The remainder of this work is structured as follows. Section~\ref{sec:related_work} discusses related work considering FFT libraries and HPX. Section~\ref{sec:methods} introduces the discrete Fourier transform and a task-level parallelization approach for two-dimensional FFT. In Section~\ref{sec:framework}, we provide additional information about the software framework. Section~\ref{sec:results} presents our benchmark results on shared memory and distributed systems. In Section~\ref{sec:conclusion}, we conclude and give an outlook on future work.

\section{Related work}\label{sec:related_work}
To the best of our knowledge, all state-of-the-art distributed multidimensional FFT implementations rely on explicit message passing with MPI.  
A recent performance comparison can be found in~\cite{Ayala2022}. The authors compare multiple FFT libraries including a handful of libraries with distributed GPU computing support. Most of the FFT libraries use FFTW~\cite{Frigo2005} as a backend. We want to include two notable mentions. First, P3DFFT~\cite{Pekurovsky2012} uses FFTW as a backend. P3DFFT is one of the first libraries supporting a pencil-based domain decomposition for three-dimensional FFTs. In contrast, plain FFTW only supports slab decomposition, which is an expansion to three dimensions of the decomposition discussed in this work. The main advantage of pencil-based decomposition is that synchronization is exclusive to row or column-wise communicators. Second, we want to highlight AccFFT~\cite{Gholami2015}. AccFFT has a cuFFT backend supporting distributed GPU computing. In~\cite{Ayala2022}, it is the best-performing library on CPU nodes and dominates the GPU benchmarks on Summit's Volta 100 accelerators. 
An overview of different asynchronous many-task systems is given in~\cite{Thoman2018}. We focus on HPX because of its conformity with the C\texttt{++} standard~\cite{hpx}. HPX can target various hardware via different executor backends~\cite{Daiss2023}. Furthermore, HPX supports multiple parcelports that enable distributed computing. A recent addition is the Lightweight Communication Interface (LCI) parcelport \cite{Yan23}. Regarding tasking overheads, HPX, Charm\texttt{++}, and \textit{MPI+OpenMP} were quantified in~\cite{wu2022quantifying}. HPX currently lacks a versatile application portfolio. However, the stellar science simulation code Octo-Tiger is based on HPX~\cite{marcello2021octo} and can be considered the flagship application. Octo-Tiger implements a fast multipole method based on octrees.

\section{Methods}\label{sec:methods}
This section introduces basic Fourier transform theory and discusses how the multidimensional FFT algorithm can be parallelized. The last subsection presents different implementation variants to determine performance differences and overheads within HPX. Furthermore, we state the benchmark hardware.\newline
We specifically focus on real-to-complex FFT in this work. There are other complex-to-complex or real-to-real transformations. However, only the basis function used by the FFT backend differs, while other aspects, such as synchronization and communication, are equivalent.

\subsection{Fast Fourier Transform}
The Fourier transform of a one-dimensional real-valued discrete signal\newline $f=[f_0,....,f_{N-1}]$ is given by 
\begin{equation}
    \hat{f}_k = \sum^{N-1}_{n=0} f_n \cdot \phi_k(n)   
\end{equation}
for $k=0, ...,\frac{N}{2}$ with $\phi_k(n) = \exp(-2\pi i \cdot \frac{n\cdot k}{N})$ and imaginary unit $i=\sqrt{-1}$. A naive algorithm can compute $\hat{f}$ in $\mathcal{O}(N^2)$. Cooley and Tukey proposed a divide and conquer algorithm to compute the transform in $\mathcal{O}(N\cdot \log(N))$ for signal lengths to the power of two~\cite{Cooley1965}. Over time, the FFT algorithm was improved, \emph{e.g.}, by using different radices~\cite{burrus1991}. The amount of multiplications is reduced at the cost of more additions and larger intermediate storage. We refer to~\cite{burrus1991} for a detailed comparison of different algorithms. 

The Fourier transform can be easily expanded to multiple dimensions. 
In practice, $\hat{F}$ can be computed by subsequent FFTs along each dimension. Note that the FFTs along the first dimension transform the signal into a complex signal. As a result, the second dimension requires complex-to-complex FFTs.

\subsection{Parallelization}\label{parallel}
We first consider shared-memory parallelism.
For task futurization, it is essential to determine the dependencies of tasks. Here, we define different tasks in the two-dimensional FFT algorithm. Furthermore, we show how they depend on each other and optimize synchronization. We assume a signal matrix of size $N \times M$ stored in row-major format. To ensure that the data for the one-dimensional FFTs is contiguous in memory, a basic algorithm consists of four computation steps:
\begin{itemize}
    \item Real-to-complex one-dimensional FFTs in the first dimension
    \item Transpose matrix for the second dimension to be contiguous in memory
    \item Complex-to-complex one-dimensional FFTs in the second dimension
    \item Transpose matrix back to the original data layout
\end{itemize}

\subsubsection{One-dimensional FFT}
As we use a backend to compute the one-dimensional FFTs, the smallest task size is given by the computation of one FFT. The task size can be enlarged by packing multiple FFTs that are contiguous in memory. For small matrices, one FFT can be too small to hide the launch overhead of the task. Here, an adjustable task size can improve performance. 

\subsubsection{Transpose}
With a sequential algorithm, there are two approaches to compute the one-dimensional FFTs in the second dimension. First, it is possible to keep the data layout with the first dimension contiguous in memory and use a strided access scheme. Therefore, the next element is offset with $N$. This approach requires less memory and performs better than an explicit transpose of a small matrix. Nevertheless, transposing becomes more performant when the matrix rows no longer fit in the cache.

Defining the smallest task size is related to the FFT task size. After an FFT task is completed, a transpose task can, independent of the progress of other FFT tasks, insert row elements into the non-contiguous columns of the transposed matrix.
All transpose tasks must finish for the second dimension FFT tasks to start. A global synchronization barrier is required. While this approach is intuitive, its performance is sub-optimal. Shifting the global synchronization barrier before the transpose tasks makes it possible to define tasks that do not read but write elements into contiguous memory.
The impact on the algorithm performance is crucial for large matrices, as described in Section~\ref{sec:results}. Note that the performance is optimized not by softening or removing the synchronization barrier. Instead, we choose the task dependencies in a more memory-oriented fashion.
\newline\newline
In the distributed two-dimensional FFT algorithm, data transfer between compute hosts or, in HPX terms, localities is necessary at two points. Once between the FFTs in the first and second dimension and once after the FFTs in the second dimension to get back to the original data layout on all $N_{locs}$ localities. This adds more steps to the algorithm compared to the shared memory version. One is the communication step, and the other is a data layout rearrangement step. The question is whether to transpose before or after the communication steps.
Since the communication step requires a global barrier beforehand, we do the transpose afterward.

\subsubsection{Communicate}
The communication scheme for distributed two-dimensional FFT is straightforward. Only $\frac{1}{N_{locs}}$ of the partial data remains on its current locality while the rest is sent to all other localities. Many distributed FFT implementations simplify communication by relying on MPI collectives.
In contrast to message-based communication libraries, \emph{e.g.}, Open MPI, where the idea is to send data to nodes to do the work, HPX aims to schedule tasks to the localities that hold the data to minimize data transfer. While this approach is rather elegant, it is not ideal for multidimensional FFT and applications with high data transfer. We know beforehand that $(1-\frac{1}{N_{locs}})$ of the data has to be transferred to other localities. Relying on the implicit communication HPX allows, with its active global address space (AGAS), does not make sense. Instead, we use the HPX equivalents of the MPI collective operations to transfer the data explicitly.
\subsubsection{Rearrange}
The rearrange tasks perform the same amount of memory copies independent from when data is transposed. The tasks either split the local matrix into equally sized $N_{locs}$ parts or concatenate the $N_{locs}$ parts collected from other localities.

\subsection{Different implementations}\label{sec:implementations}
We designed several implementations of the same two-dimensional real-to-complex FFT algorithm to evaluate potential overheads within HPX:
\begin{itemize}
    \item \textit{HPX future naive}: Intuitive future-based implementation that defines tasks to postpone or remove synchronization points.
    \item \textit{HPX future opt}: Future-based implementation with optimized transpose.
    \item \textit{HPX future sync}: Future-based implementation with synchronization after each algorithmic step.
    \item \textit{HPX future agas}: Future-based implementation with (redundant) AGAS.
    \item \textit{HPX for\_loop}: Implementation using a synchronizing HPX feature called \lstinline{hpx::experimental::for_loop}.
    \item \textit{HPX future agas dist}: Future-based distributed implementation with AGAS and scatter or all\_to\_all collectives.
    \item \textit{HPX for\_loop dist}: Distributed implementation using the synchronizing HPX feature \lstinline{hpx::experimental::for_loop}, scatter,and all\_to\_all collectives.
\end{itemize}
Two hardware system setups are used in this work. 
The distributed benchmark is computed on a $16$-node cluster with homogeneous nodes. Each node contains a dual-socket AMD EPYC $7352$\footnote{\url{https://www.amd.com/de/products/cpu/amd-epyc-7352} (visited on 01/15/2024)} with $48$ physical cores and a total of $256$ MB of L3 cache. For our shared memory comparisons, we switch to a dual-socket AMD EPYC $7742$\footnote{\url{https://www.amd.com/de/products/cpu/amd-epyc-7742} (visited on 01/15/2024)} machine with $128$ physical cores and a combined L3 cache of $512$ MB to examine more scaling beyond 100 cores.

\section{Software framework}\label{sec:framework}
The two principal software tools we use are the C\texttt{++} standard library for parallelism and concurrency HPX~\cite{hpx} and the FFT library FFTW~\cite{Frigo2005}.

\subsection{HPX}
HPX is an asynchronous, many-task runtime system implementing the ISO C\texttt{++} language's support for data parallelism. HPX offers fine-grained parallelism and lightweight synchronization using futures and local control objects (latches, barriers, etc). The features provided in HPX promote task-based algorithm decomposition. 
In addition, HPX implements an AGAS, allowing users to construct distributed data types and structures. Users can deploy functions and methods on the distributed data types all by using the future and local control objects to manage parallelism and concurrency. With the parcelport concept, HPX separates communication from a specific standard. Several parcelports are supported, ranging from TCP and MPI to the recently added LCI parcelport \cite{Yan23}. 

HPX contains several high-level parallelization features. We use the \newline\lstinline{hpx::experimental::for_loop} in this work. The syntax is similar to a classic C\texttt{++} loop. However, it allows for the specification of a parallel execution policy similar to \lstinline{std::for_each}. To optimize performance, the iterations are bundled into tasks. At the end of the loop, all tasks are synchronized.

\subsection{FFTW}
Since $20$ years, the de-facto standard has been the Fastest Fourier Transform in the West (FFTW) library~\cite{Frigo2005}. To date, FFTW is a very performant FFT library, which still serves as a  backend for many state-of-the-art libraries.
The secret of FFTW's performance lies in planning. Each FFT requires a plan. Depending on the available resources, several planning options can be set. Estimated planning minimizes a heuristic cost function based on FLOPS and memory accesses. 
In contrast, patient planning uses dynamic programming to find an efficient algorithm from a set of plan combinations. A middle ground is measured (formerly impatient) planning that works with a reduced algorithm set. FFTW requires runtime initialization, a function to distribute work to the runtime system's workload manager, and runtime finalization.
FFTW supports parallelization by offering several backends. Distributed FFTs are supported via an MPI backend. For shared memory parallelization, FFTW has a threads interface that supports pthreads~\cite{Nichols1996} and OpenMP~\cite{Chandra2001} out of the box.
We add an HPX backend to FFTW. The HPX backend supports all functional requirements of the FFTW threads interface.
A more detailed explanation of the process can be found in Subsection \ref{sec:fftw_backend}.

Furthermore, we use FFTW as the one-dimensional backend for our HPX-based parallelization. All tasks can reuse FFTW plans as planning is only tied to the FFT length. Bringing multiple one-dimensional FFTs contiguous in memory is possible in one plan. 

\section{Results}\label{sec:results}
We divide the results section into three parts. First, we consider a shared memory benchmark to compare our HPX-based implementations and expose overheads induced by code structure and HPX features. Second, we benchmark our best implementation against FFTW's backends, including the new HPX backend. The benchmark takes into consideration the impact of planning time. Third, we turn our attention to a distributed benchmark and compare the \textit{MPI+X} approach of FFTW to purely HPX-based implementations based on collective operations. We use synthetic data and a problem size of $2^{14} \times 2^{14}$. For all runtimes, we take the median out of 50 runs and visualize error bars related to the best and worst runs. For additional information about the reproducibility of our results, we refer to the supplementary material.

\subsection{Overheads}\label{sec:overheads}
\begin{figure}
    \centering
    \begin{minipage}[t]{.47\textwidth}
        \centering
        \includegraphics[width=\linewidth]{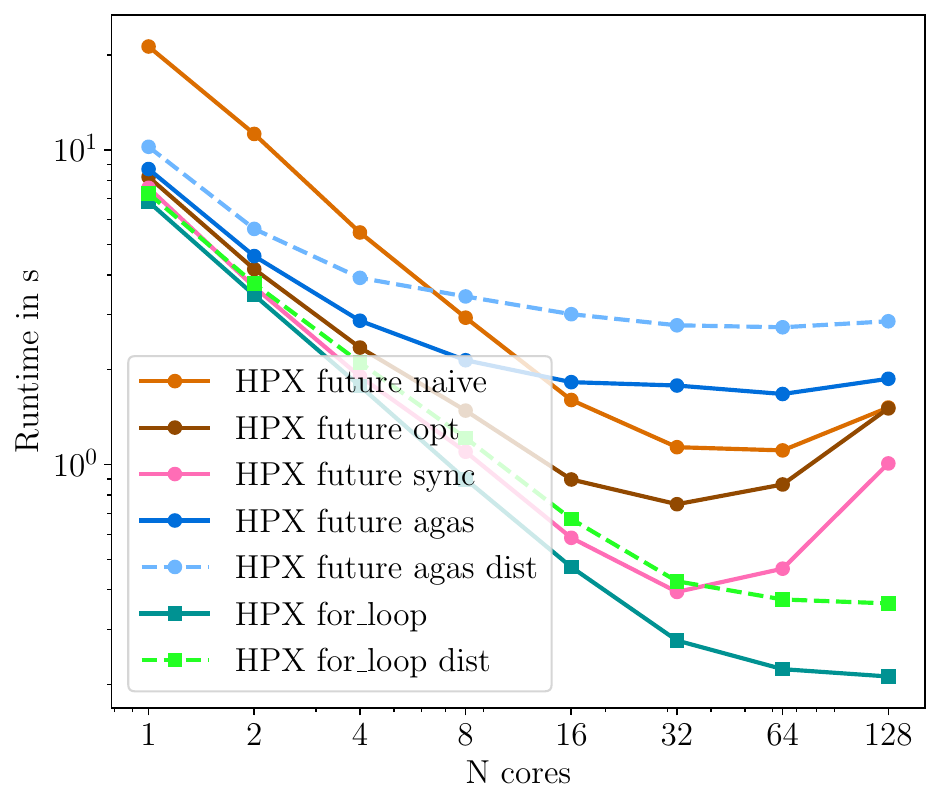}
        \caption{Strong scaling runtimes on up to $128$ cores for a $2^{14} \times 2^{14}$ FFT. Task-futurization-based implementations are visualized with circles, while implementations based on \lstinline{hpx::experimental::for_loop} are visualized with squares. Implementations capable of distributed computing are highlighted with dashed lines.}
        \label{fig:shared_runtime_16384_hpx}
    \end{minipage}\hspace{.05\textwidth}
    \begin{minipage}[t]{.47\textwidth}
        \centering
        \includegraphics[width=\linewidth]{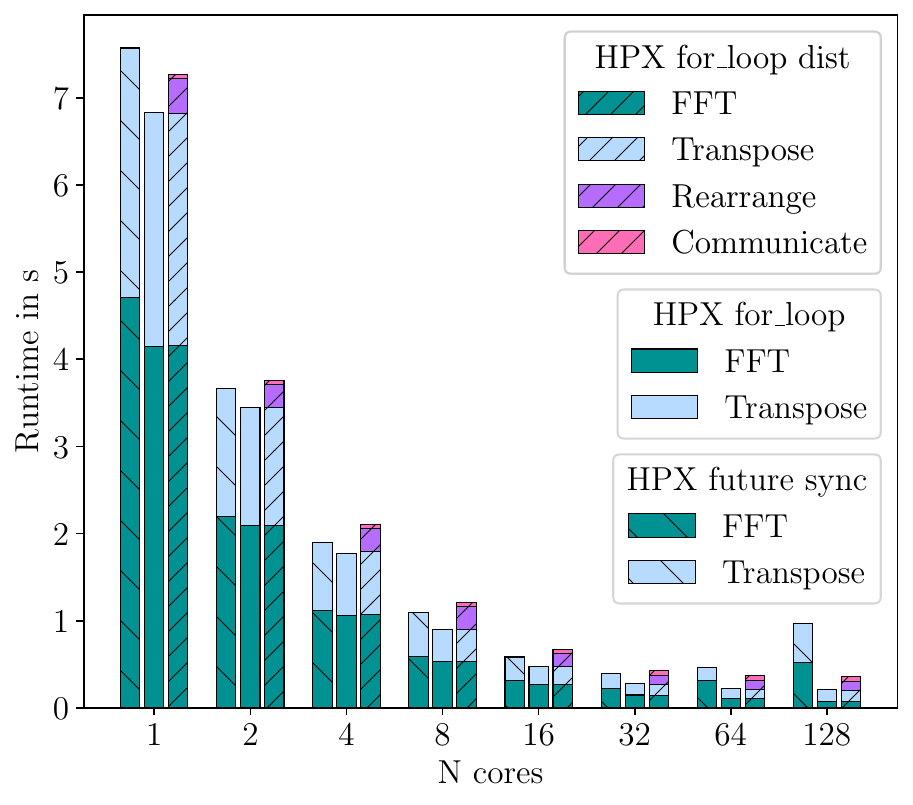}
        \caption{Selected strong scaling runtime distributions on up to $128$ cores for a $2^{14} \times 2^{14}$ FFT (compare Figure~\ref{fig:shared_runtime_16384_hpx}). The partial runtimes correspond to the computation steps in Section~\ref{parallel}.} 
        \label{fig:shared_distribution_16384}
    \end{minipage}
\end{figure}

We use the shared memory system described in Section~\ref{sec:implementations} to compare our different implementations. The strong scaling results on up to $128$ cores are illustrated in Figure~\ref{fig:shared_runtime_16384_hpx}. For the one-dimensional FFTs, we use estimated planning within the FFTW backend, resulting in neglectable planning time. To interpret the result correctly, it is essential to consider the system's hardware architecture. It contains two sockets with $64$-core CPUs based on a chiplet design. Each CPU contains eight CCDs connected via AMD Infinity Fabric over an IO die.

First, observe the naive (orange) and optimized (brown) variants. Shifting the synchronization barrier such that cache performance is optimized has a visible impact on performance. Moreover, the completely synchronized future-based implementation (pink) performs better than its asynchronous counterparts. Computing all transpose tasks at the same time further optimizes cache usage. These results perfectly visualize a common pitfall when applications are ported to asynchronous tasks. Reducing task dependencies and removing global synchronization barriers does not necessarily improve performance. When defining task dependencies, not only the algorithm but also other critical performance indicators, \emph{e.g.}, cache performance, must be considered. The variant that accesses actions through the AGAS (dark blue) is plotted to visualize the induced overhead. The task size is too small to hide the overhead.
The implementation based on the \lstinline{hpx::experimental::for_loop} (pine) performs better than all future-based variants. It enforces task synchronization after the loop, making it the perfect choice for the FFT application.
The redundant communication overhead of distributed implementations is visualized with dashed lines.

Figure~\ref{fig:shared_distribution_16384} shows the runtime decomposition for selected synchronized versions. Compared to the \lstinline{hpx::experimental::for_loop} variants, the future-based variant has more extensive runtimes across the board. These results further highlight the efficiency of the \lstinline{hpx::experimental::for_loop}. The communication overhead is constant as a collective operation simplifies to a concatenation of move operations. 

\subsection{FFTW Backend}\label{sec:fftw_backend}
Apart from analyzing overheads within HPX, providing an HPX backend for FFTW is a major contribution to this work\footnote{\url{https://github.com/FFTW/fftw3/pull/341} (visited on 02/04/2024)}. The HPX backend implements the existing threading interface provided by FFTW. The HPX backend is an optional compile-time target for FFTW. To start an initial HPX thread, the HPX backend uses \lstinline{hpx::threads::run_as_hpx_thread}. The initial HPX thread segments the dataset and starts the asynchronous execution of threads on each segment using \lstinline{hpx::async}.
One of the challenges the HPX backend faced was the proper forwarding of runtime system arguments. 
The HPX backend solves this challenge by adding an environment variable, \lstinline[language=bash]{FFTW_HPX_NTHREADS}. The new environment variable allows users to set the number of OS-level threads HPX creates during program and runtime system initialization. Without the environment variable, HPX will create a number of OS-level threads equal to the number of available cores returned by \textit{hwloc}. The HPX default thread creation behavior is appropriate as a general solution, the new environment variable makes the performance analysis in this paper possible.

\begin{figure}
    \centering
    \begin{minipage}[t]{.47\textwidth}
        \centering
        \includegraphics[width=\linewidth]{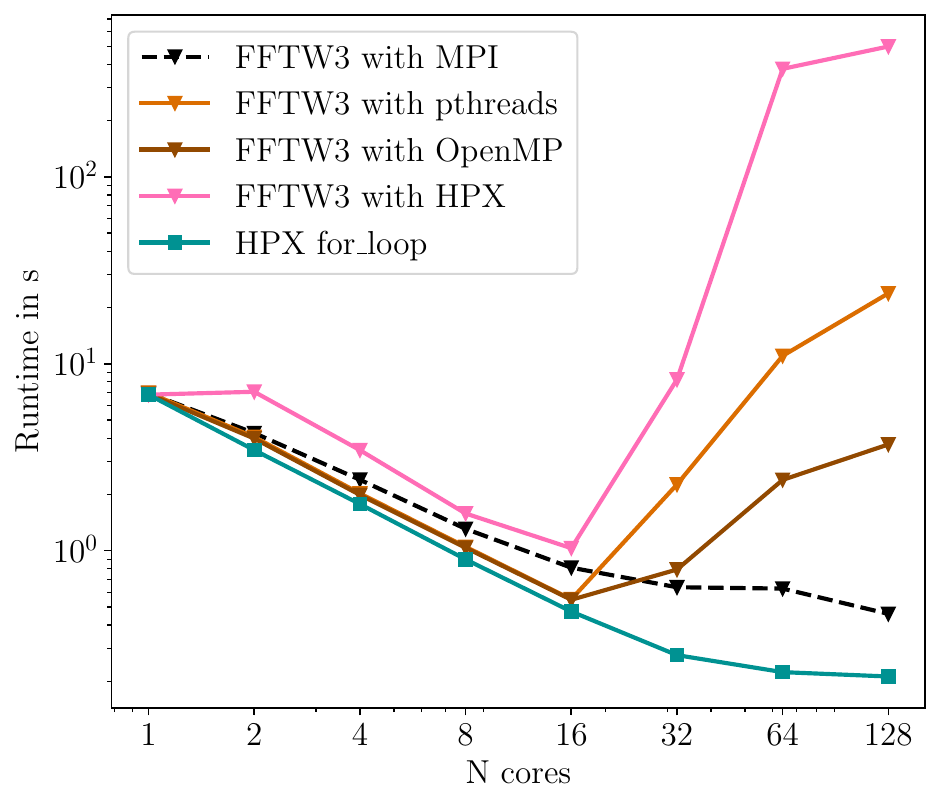}
    \end{minipage}\hspace{.05\textwidth}
    \begin{minipage}[t]{.47\textwidth}
        \centering
        \includegraphics[width=\linewidth]{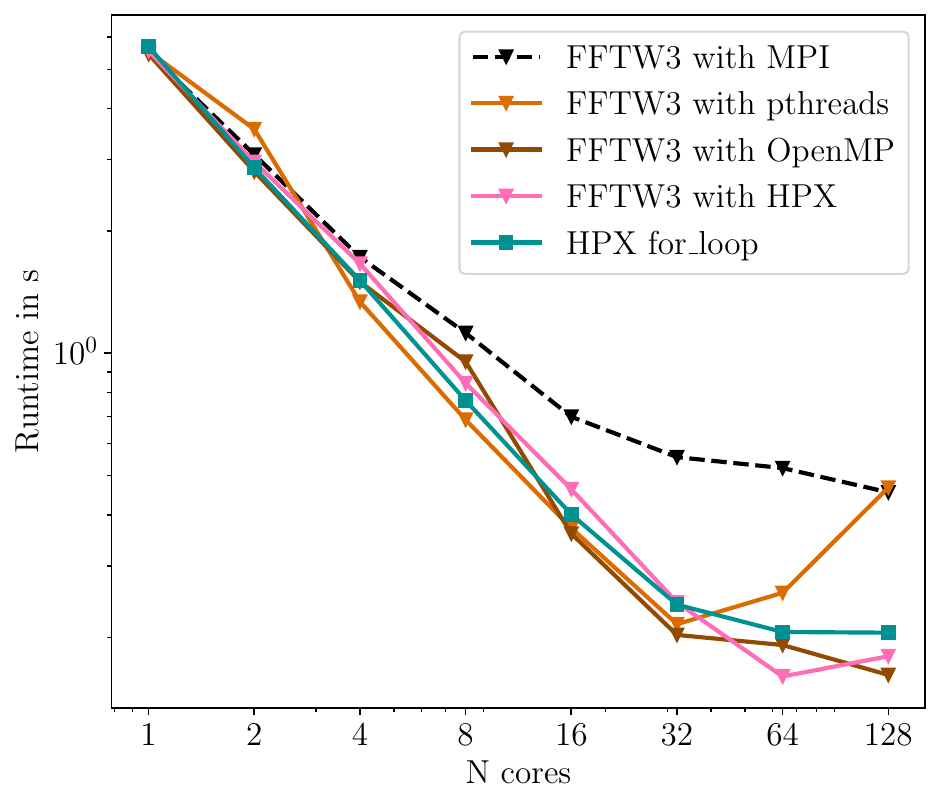}
    \end{minipage}
    \caption{Strong scaling runtimes on up to $128$ cores for a $2^{14} \times 2^{14}$ FFT. The different FFTW's backends are visualized with triangles. FFTW's ESTIMATE planning (\textbf{left}) is compared to MEASURE planning (\textbf{right}).
    The runtimes of the \lstinline{hpx::experimental::for_loop} implementation using the respective planning are also included.} 
    \label{fig:shared_runtime_16384_fftw3}
\end{figure}

Figure~\ref{fig:shared_runtime_16384_fftw3} shows the HPX backend's performance compared to existing FFTW backends using estimated and measured planning. The runtime of our own HPX parallelization is added for scale. The HPX backend performs consistently worse than the other backends with FFTW's estimated planning. The runtimes skyrocket for all three threading backends, while the runtime of the MPI backend does not. According to our investigation, this is related to FFTW planning. Overall, our HPX-based parallelization has the best performance and scaling.

We switch to measured planning to increase the threading backends' performance for more than 16 cores. While the MPI backend does only marginally profit, the effect on the threading backends is enormous. The HPX backend now rivals the performance of the existing backends. In addition, our \lstinline{hpx::experimental::for_loop} implementation also marginally profits from measured planning. 

The corresponding planning times can be observed in Figure~\ref{fig:shared_runtime_16384_fftw3_measure_plan}. Since our parallelization only requires two one-dimensional plans, the planning time is not dependent on the core count. Two-dimensional planning is more complicated, requiring strided memory access or data transposition. Thus, the planning time is over factor $50$ slower averaged over all FFTW backends. 
The OpenMP and pthreads backends require nearly the same amount of planning time. In contrast, FFTW with the HPX backend requires more time to create a plan. This is likely caused by HPX being poorly suited for some plans in the planning set, \emph{e.g.}, plans similar to the estimated plan.

\subsection{Distributed}
\begin{figure}
    \centering
    \begin{minipage}[t]{.47\textwidth}
        \centering
        \includegraphics[width=\linewidth]{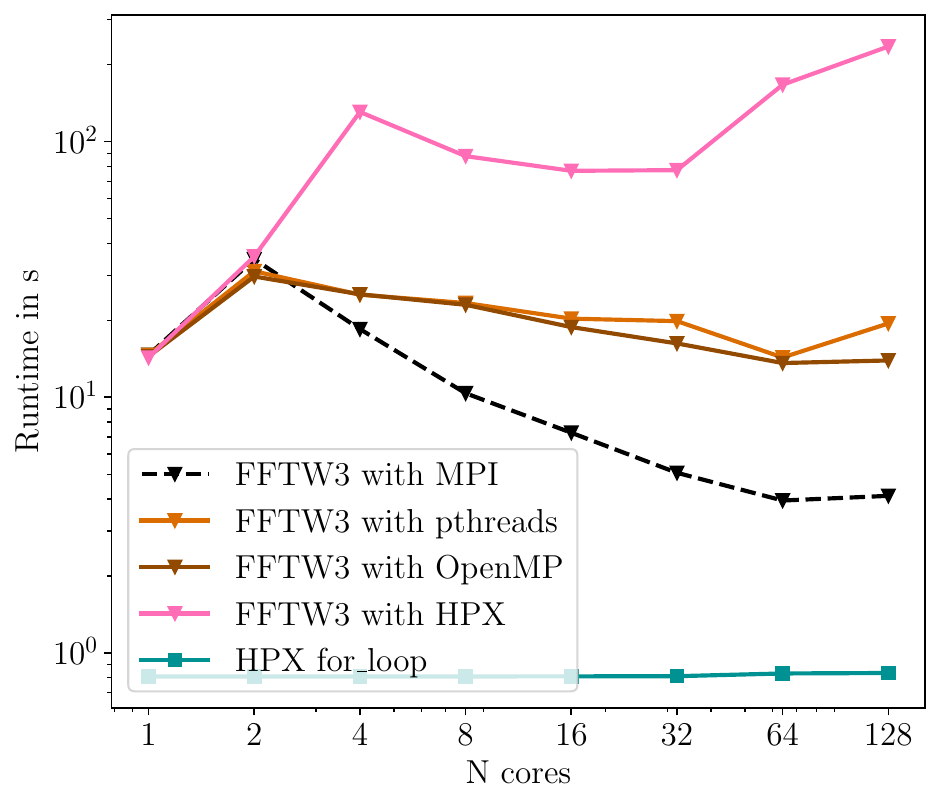}
        \caption{Measured planing time for FFTW backend strong scaling (see Figure~\ref{fig:shared_runtime_16384_fftw3}). The different FFTW's backends are visualized with triangles. The plan times of the \lstinline{hpx::experimental::for_loop} implementation are included for comparison.} 
        \label{fig:shared_runtime_16384_fftw3_measure_plan}
    \end{minipage}\hspace{.05\textwidth}
    \begin{minipage}[t]{.47\textwidth}
        \centering
        \includegraphics[width=\linewidth]{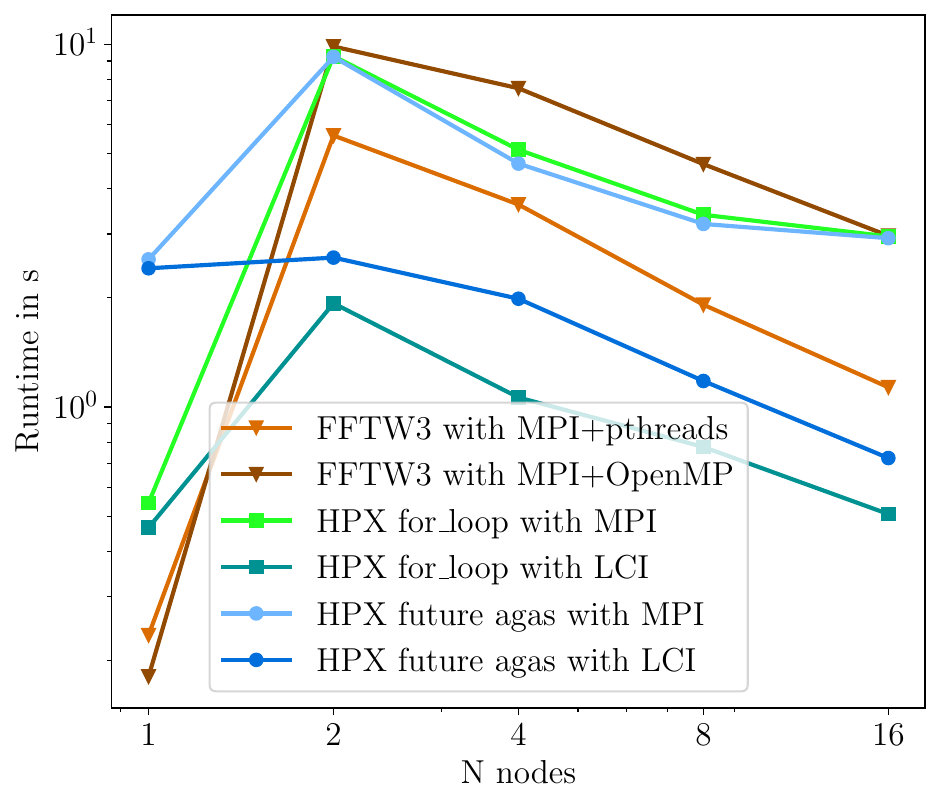}
        \caption{Distributed strong scaling on a 16-node cluster with $48$-core nodes for a $2^{14} \times2^{14}$ matrix. The FFTW backends are visualized with triangles, while our HPX implementations are visualized using circles and squares (compare Figure~\ref{fig:shared_runtime_16384_hpx}.} 
        \label{fig:dist_runtime}
    \end{minipage}
\end{figure}

We benchmark our distributed implementations based on the HPX collective operations and compare them against FFTW's parallelization based on \textit{MPI+X} (see Figure~\ref{fig:dist_runtime}). 
We set the number of threads of the shared memory parallelization to $48$. Note that while the problem size is small for a cluster, it allows highlighting the communication overheads of the MPI and LCI parcelports. The results on one compute host match the result from Section~\ref{sec:overheads} and \ref{sec:fftw_backend}. All implementations perform worse in a distributed setting due to the extensive communication overhead. The HPX collectives using the MPI parcelport can compete with FFTW's \textit{MPI+OpenMP} but get outperformed by \textit{MPI+pthreads}. Nonetheless, our pure HPX implementations taking advantage of the LCI parcelport have the best performance. Compared to the MPI parcelport we observe a communication speed-up of factor $4$ to $5$. In comparison to FFTW with \textit{MPI+pthreads}, we observe a speed-up of factor $1.5$ to $2$.
As a threading backend, FFTW currently supports OpenMP and pthreads. Providing an \textit{MPI+HPX} backend is challenging since MPI would run on the OS system threads and HPX on its lightweight threads. The different threads interfere with each other. For that reason, the HPX backend for FFTW does not support MPI+HPX.
        
\section{Conclusion and outlook}\label{sec:conclusion}
In the scope of this work, we presented several different parallel implementations of a multidimensional FFT. We quantified overheads and revealed common misconceptions when porting code to asynchronous tasks. Asynchronous algorithms are not more efficient by default and can even perform worse. Cache performance is crucial when designing task graphs of parallel algorithms. For multidimensional FFT, the \lstinline{hpx::experimental::for_loop} proved the most effective HPX tool. It yielded the fastest runtimes and the best scaling.

Furthermore, we implemented a shared memory backend for FFTW that uses HPX threading. 
A performance evaluation of the new FFTW HPX backend, compared to the existing FFTW backends, is made. Although FFTW's parallelization is not optimized for asynchronous runtime systems, the HPX backend performance is competitive with the OpenMP backend. Both show excellent scaling if FFTW uses measured planning. However, the HPX backend requires nearly ten times as much planning time. This can become an issue if plans are computed online and only executed a few times. Our parallelization using the \lstinline{hpx::experimental::for_loop} is competitive with all FFTW backends while requiring significantly shorter planning time.
Furthermore, we highlighted the communication overhead and compared FFTW's \textit{MPI+X} to HPX collectives regarding distributed performance. While FFTW's \textit{MPI+OpenMP} performed comparably to the HPX collectives using the MPI parcelport, the LCI parcelport resulted in a speed-up of up to factor $5$ on our benchmark cluster. 

Considering future work, we plan to expand our distributed testing and evaluate distributed HPX support for FFTW. In addition, we plan to run benchmarks on more hardware architectures, e.g., ARM or RISC-V, focusing on the power efficiency and performance portability of the different FFTW backends and our HPX parallelization.

\section*{Supplementary materials}
The source code and benchmark scripts are available at \href{https://doi.org/10.18419/darus-4094}{DaRUS}\footnote{\url{https://doi.org/10.18419/darus-4094} (visited on 13/03/2024)}. The respective software and compiler versions are stated in the \lstinline[language=bash]{README.md} of the source code. 

\bibliographystyle{splncs04}
\bibliography{bibliography}

\begin{thebibliography}{10}
\providecommand{\url}[1]{\texttt{#1}}
\providecommand{\urlprefix}{URL }
\providecommand{\doi}[1]{https://doi.org/#1}

\bibitem{Ayala2022}
Ayala, A., et. al.: {FFT} benchmark performance experimentson systems targeting exascale. Tech. rep., University of Tennessee (2022)

\bibitem{burrus1991}
Burrus, C.S., Parks, T.W.: DFT/FFT and Convolution Algorithms: Theory and Implementation. John Wiley \& Sons, Inc., USA, 1st edn. (1991)

\bibitem{Chandra2001}
Chandra, R., Dagum, L., Kohr, D., Menon, R., Maydan, D., McDonald, J.: Parallel programming in OpenMP. Morgan kaufmann (2001)

\bibitem{Cooley1965}
Cooley, J., Tukey, J.: An algorithm for the machine calculation of complex fourier series. Mathematics of Computation  \textbf{19}(90),  297--301 (1965)

\bibitem{Daiss2023}
Dai\ss{}, G., et. al.: Stellar mergers with {HPX-Kokkos} and {SYCL}: Methods of using an asynchronous many-task runtime system with {SYCL}. In: IWOCL '23. ACM, New York, NY, USA (2023)

\bibitem{Deserno1998}
Deserno, M., Holm, C.: {How to mesh up Ewald sums. I. A theoretical and numerical comparison of various particle mesh routines}. J. Chem. Phys.  \textbf{109}(18),  7678--7693 (11 1998)

\bibitem{Frigo2005}
Frigo, M., Johnson, S.: {The Design and Implementation of FFTW3}. Proceedings of the IEEE  \textbf{93}(2),  216--231 (2005)

\bibitem{Gabriel2004}
Gabriel, E., et. al.: Open {MPI}: Goals, concept, and design of a next generation {MPI} implementation. In: Proceedings, 11th European PVM/MPI Users' Group Meeting. pp. 97--104. Budapest, Hungary (September 2004)

\bibitem{Gholami2015}
Gholami, A., et. al.: Accfft: {A} library for distributed-memory {FFT} on {CPU} and {GPU} architectures. CoRR  (2015)

\bibitem{hpx}
Kaiser, H., et~al.: {HPX} -- the {C\texttt{++}} standard library for parallelism and concurrency. Journal of Open Source Software  \textbf{5}(53), ~2352 (2020)

\bibitem{Lavin2016}
Lavin, A., Gray, S.: Fast algorithms for convolutional neural networks. In: CVPR. pp. 4013--4021 (June 2016)

\bibitem{marcello2021octo}
Marcello, D.C., et. al.: {Octo-Tiger: a new, 3D hydrodynamic code for stellar mergers that uses HPX parallelization}. MNRAS  \textbf{504}(4),  5345--5382 (2021)

\bibitem{Nichols1996}
Nichols, B., Buttlar, D., Farrell, J.P.: Pthreads Programming. O'Reilly \& Associates, Inc., USA (1996)

\bibitem{Pekurovsky2012}
Pekurovsky, D.: P3dfft: A framework for parallel computations of fourier transforms in three dimensions. SISC  \textbf{34}(4),  C192--C209 (2012)

\bibitem{Thoman2018}
Thoman, P., et. al.: A taxonomy of task-based parallel programming technologies for high-performance computing. J. Supercomput.  \textbf{74}(4),  1422–1434 (04 2018)

\bibitem{Wallace1991}
Wallace, G.K.: {The JPEG Still Picture Compression Standard}. Commun. ACM  \textbf{34}(4),  30–44 (apr 1991)

\bibitem{wu2022quantifying}
Wu, N., et. al.: {Quantifying Overheads in Charm\texttt{++} and HPX Using Task Bench}. In: Euro-Par. pp. 5--16. Springer (2022)

\bibitem{Yan23}
Yan, J., Kaiser, H., Snir, M.: {Design and Analysis of the Network Software Stack of an Asynchronous Many-task System -- The LCI parcelport of HPX}. In: Proceedings of the SC '23 Workshops. p. 1151–1161. ACM, New York, NY, USA (2023)

\end{thebibliography}
\end{document}